\documentclass {article}

\usepackage{PRIMEarxiv}

\usepackage[utf8]{inputenc} 
\usepackage[T1]{fontenc}    
\usepackage{booktabs}       
\usepackage{amsfonts}       
\usepackage{nicefrac}       
\usepackage{microtype}      
\usepackage{lipsum}
\usepackage{fancyhdr}       
\usepackage{graphicx}       
\graphicspath{{media/}}     
\usepackage{amsmath}
\usepackage{soul}
\usepackage{graphicx}
\usepackage[hyphens,spaces,obeyspaces]{url}
\usepackage{xfrac}
\usepackage{booktabs}
\usepackage{algorithm}
 \usepackage[]{algpseudocode}
\usepackage{tikz}
\usetikzlibrary{arrows,shapes,snakes,automata,backgrounds,petri,fit,positioning, chains}
\usepackage{eqparbox}
\usepackage{url}
\usepackage{hyperref}
\usepackage{natbib}

\algdef{SE}[SUBALG]{Indent}{EndIndent}{}{\algorithmicend\ }%
\algtext*{Indent}
\algtext*{EndIndent}
 
\usepackage{color}
\usepackage{soul}

\usepackage{subfig}
\usepackage{color, colortbl}
\definecolor{Gray}{gray}{0.9}
\newcommand\correspondingauthor{\thanks{Corresponding author.}}

\usepackage{mwe}
\usepackage{tikz}
\title{A CRISP-DM-based Methodology for Assessing Agent-based Simulation Models using Process Mining
}

\author{
  Rob H. Bemthuis \correspondingauthor \\
  University of Twente, Enschede, Netherlands \\
  Faculty of Electrical Engineering, \\Mathematics and Computer Science\\
  \texttt{r.h.bemthuis@utwente.nl} \\
   \And
  Ruben R. Govers\\
  University of Twente, Enschede, Netherlands \\
  Faculty of Electrical Engineering, \\Mathematics and Computer Science\\
  \texttt{r.r.govers@student.utwente.nl} \\
  \AND
   Amin Asadi \\
  University of Twente, Enschede, Netherlands \\
  Department of High-tech Business and\\ Entrepreneurship\\
  \texttt{amin.asadi@utwente.nl} \\
}

\begin{document}
\maketitle

\begin{abstract}
Agent-based simulation (ABS) models are potent tools for analyzing complex systems. However, understanding and validating ABS models can be a significant challenge. To address this challenge, cutting-edge data-driven techniques offer sophisticated capabilities for analyzing the outcomes of ABS models. One such technique is process mining, which encompasses a range of methods for discovering, monitoring, and enhancing processes by extracting knowledge from event logs. However, applying process mining to event logs derived from ABSs is not trivial, and deriving meaningful insights from the resulting process models adds an additional layer of complexity. Although process mining is invaluable in extracting insights from ABS models, there is a lack of comprehensive methodological guidance for its application in ABS evaluation in the research landscape. In this paper, we propose a methodology, based on the CRoss-Industry Standard Process for Data Mining (CRISP-DM) methodology, to assess ABS models using process mining techniques. We incorporate process mining techniques into the stages of the CRISP-DM methodology, facilitating the analysis of ABS model behaviors and their underlying processes. We demonstrate our methodology using an established agent-based model, Schelling’s model of segregation. Our results show that our proposed methodology can effectively assess ABS models through produced event logs, potentially paving the way for enhanced agent-based model validity and more insightful decision-making. 
\end{abstract}


\section{Introduction}

\label{section:introduction}
Agent-based simulation (ABS) models have become increasingly popular in a variety of fields, including construction~\citep{khodabandelu2021agent}, energy systems~\citep{hansen2019agent,castro2020review}, transportation~\citep{mahmud2021hybrid,fehn2023integrating,mei2023multi}, and biology~\citep{zhang2020overview}. These models provide a powerful tool for modeling and analyzing complex socio-technical systems by simulating the behavior of individual agents within a system and their interactions with each other and the environment. ABS models are particularly adept at representing complex systems that are difficult to characterize using traditional analytical methods. Recent advancements in computational techniques, algorithms, and data-driven methods have led to the development of increasingly advanced ABS models~\citep{abar2017agent,deangelis2019decision}. However, as the complexity of these systems continues to grow, there is a growing need for sophisticated models that can capture the intricacies of the interactions between their components~\citep{troost2023keep,ouda2023comprehensive}. 

The success of ABS models depends on their ability to accurately represent reality. Creating valid and verifiable ABS models is challenging due to several factors, including the lack of established standard verification methods~\citep{gurcan2013generic,xiang2005verification}, the difficulty in defining appropriate evaluation metrics~\citep{liu2005study}, the inherent complexity exhibited by these models~\citep{Benjamin2014,fagiolo2019validation}, and the need for human expert knowledge to accurately model them~\citep{troost2023keep}. Another consideration is that ABS models typically target system-level outputs that emerge from micro-level agent behaviors, which are governed by basic rules~\citep{taveter2001agent,bonabeau2002agent,macal2009agent}. However, these emergent behaviors can be unexpected, counter-intuitive, and challenging to understand, if they can be captured at all. 

Additionally, the multi-tiered nature of ABS models, with individual agents operating at a lower level and larger-scale patterns emerging at a higher level, calls for a multi-level approach for thorough analysis~\citep{parry2011large,brugiere2022handling}, also in the context of validation~\citep{chen2008method,soyez2013methodology}. This could entail not only comparing the input-output relationships of the entire system but also implementing analysis procedures for agent sub-groups or partial models, extending to individual agents~\citep{kasaie2015guidelines}. Nevertheless, ensuring comprehensive validation typically demands significant time and effort. In real-world scenarios, it is key to have access to data that can validate model behavior across all levels. 

Advances in data-driven techniques provide novel avenues for system behavior analysis, enhancing the precision and credibility of ABS models~\citep{baqueiro2009integrating,angione2022using,wang2022preparing,patsatzis2023data}. One such data-driven discipline is process mining, which has emerged as a promising means for the verification and validation of ABS models~\citep{cabac2006analysis,vsperka2013control,halavska2019advantages}. Process mining encompasses a range of methods for discovering, monitoring, and improving processes by extracting knowledge from event logs~\citep{van2011process}. For example, by applying process mining techniques to the output of an ABS model, patterns and anomalies in the model’s behavior can be identified~\citep{bemthuis2023CoopIS} and compared with the model’s conceptual blueprint~\citep{bemthuis2023EDOC}. Besides, it facilitates detecting discrepancies or errors in underlying assumptions or parameters~\citep{flick2010re}. This information can be used to fortify the verification and validation processes, consequently supporting decision-making in various fields. 

The application of process mining techniques combined with ABS has proven its effectiveness, as further showcased in Section~\ref{section:related work}. However, the task of applying process mining to log files produced by ABS models presents significant challenges, as also pointed out by~\citet{denz2014process} and \citet{tour2023mining}, due to the following reasons. First, ABS models are typically complex, containing a large number of agents, each with their own behavior and interactions~\citep{macal2005tutorial}. The emergent nature of these interactions makes it difficult, and if not, impossible, to predict and assess the model’s behavior, hindering the identification of patterns and behaviors. Adding to the complexity is the adaptive and evolving nature of agents. For example, this is evident in multi-agent systems where agents employ reinforcement learning to learn and adapt, a concept known as multi-agent reinforcement learning \citep{zhang2021multi,oroojlooy2023review}. Second, the output generated by ABS models for use in process mining (e.g., event logs) can be large and noisy~\citep{tour2021agent}, making it challenging to extract meaningful information~\citep{suriadi2017event}. Third, ABS model behavior is generally stochastic. Consequently, the same model may yield different results with each run, a factor to consider during process mining-based analysis~\citep{bemthuis2022discovering} but often not considered~\citep{leemans2021stochastic}. Thus, significant progress remains to be made in defining, extracting, and transforming log files generated by agent systems for the purpose of process mining. 

We use the terms ``evaluation'' and ``assessment'' interchangeably while acknowledging their nuanced distinctions in certain contexts. When addressing the ``evaluation/assessment'' of agent-based models, our attention is primarily on the examination of agent behaviors and interactions. Our delineation emphasizes the analysis of outputs, such as event logs, generated by an agent-based system. Our focus does not encompass assessing the overall performance efficiency of a simulation model; instead, we are concerned with the emergent behaviors it manifests. Similarly, while recognizing the nuances among terms like ``agent-based system'', ``agent-based simulation'', ``agent-based model'', and ``multi-agent system'' (the latter typically emphasizes multi-agent interactions) we use them interchangeably. We refer to a computational framework where agents, defined by unique attributes and behaviors, interact with each other or their environment, resulting in emergent system-wide phenomena. 

The goal of this article is to introduce a structured methodology for assessing ABS models through the application of process mining techniques. Our methodology is rooted in the Cross-Industry Standard Process for Data Mining (CRISP-DM)~\citep{wirth2000crisp} methodology and is exemplified by Schelling's segregation model, a seminal example in the domain of ABS models. Our methodology offers: (1) a guideline for the steps to undertake when applying process mining techniques to assess event logs generated by an agent-based system; and (2) an endeavor to enhance the replicability and transparency of combined ABS and process mining research by adapting a well-established data science project methodology. 

The key contributions of this study are: (1) we propose a methodology grounded in the CRISP-DM framework for evaluating ABS models using process mining techniques; (2) we apply process mining techniques to dissect the output from an ABS model; (3) we provide a practical case study of our proposed methodology using Schelling’s segregation model. 

The remainder of this article is structured as follows. Section~\ref{section:related work} discusses related work. Section~\ref{section:methodology} presents our proposed method. Section~\ref{section:illustrative scenario} demonstrate the method's effectiveness using Schelling's model of segregation. Section~\ref{section:discussion} provides a discussion of key findings and their implications. Finally, Section~\ref{section:conclusions} concludes and outlines directions for future research. 

\section{Related work}
\label{section:related work}
Several lines of research converge at the intersection of process mining and ABS systems, such as the utilization of process mining to facilitate the verification of multi-agent systems~\citep{ou2010applying}, the exploration of agent systems through process mining methodologies~\citep{bemthuis2022discovering,tour2022agent}, and the investigation into the complexities of multi-agent interactions, such as group dynamics~\citep{rozinat2009analyzing}. Although a substantial body of work exists on the automated generation of multi-agent system models from event data~\citep{tour2021agent,tour2022agent}, our focus is on the use of data produced by ABS models through process mining techniques. Subsequently, we discuss methodological nuances present in the existing literature. 

\subsection{Specialized properties of agent-based systems}
Several scholarly articles propose specific frameworks for specialized properties or design aspects of agent-based systems. For example, \citet{fauzan2019simulation} embark on an exploration of asynchronous messaging in ABSs, specifically within the context of port container terminals. Their inquiry underscores the role of such messages in influencing dwelling times during the importation of goods. \cite{mecheraoui2020compositional} discuss an approach for conformance checking (a subfield within process mining) between Petri nets and event logs of multi-agent systems. 

\subsection{Methodological approaches in agent-based systems and process mining}
Numerous studies address the intersections of agent-based systems and process mining paradigms from methodological standpoints. For example, \citet{ito2018process} propose an architecture for multi-agent systems, elucidating its symbiotic relationship with process mining. This framework integrates process mining strategies with insights from an agent abstract architecture. \citet{bemthuis2019agent} propose an enterprise architecture on the joint use of agent-based systems and process mining techniques, with the purpose of aiding supply chain managers with decision-making. To support the validity of their architecture, they discuss a case study whereby they use the extracted event logs to analyze the performances of the agent system. \citet{tour2021agent} present an approach to automate the creation of multi-agent systems. This work focuses on extracting an agent system from system log files. It provides various perspectives for analyzing the macro-level behavior of changes on the micro-level of agents~\citep{tour2021agent}. \citet{denz2012process} presents a conceptual framework tailored for the application of process mining in the analysis and validation of multi-agent systems. Grounded in this framework, the author discerns agent-centric analytical vantage points and simulation-specific use cases. The work of \citet{bemthuis2023CoopIS} and \citet{bemthuis2023EDOC} both involve the utilization of event logs for assessing the face validity of agent-based systems, respectively presenting an approach to conducting such an assessment and identifying and examining outlier behaviors. In both works, the authors emphasize the importance of methodological approaches for analyzing ABS models using process mining techniques. \citet{nesterov2023discovering} present a compositional approach to obtain architecture-aware process models of multi-agent systems. They inspect the individual behavior of agents and discover a multi-agent system by combining the agent models discovered from the individual log files. As an intermediate model, they design a collection of interface patterns that describe typical agent interactions. 

Existing literature primarily enumerates potential analytical perspectives and applications of agents and process mining, rather than providing a concrete, methodological plan for implementing data-centric projects. While the framework proposed by \citet{denz2012process} illuminates aspects of agent-based system analysis and the depth of analysis attainable through process mining, our work goes beyond this discussion. We not only encapsulate analytical paradigms of agent and process mining but also offer practical guidelines on specific tasks that necessitate execution. 

In summary, the existing literature underscores the potential of employing process mining techniques for data produced by agent-based systems. However, there remains a need for methodological guidance to evaluate such integrative models. For instance, a uniform methodology encompassing models, programming languages, and tools is, to the best of our knowledge, absent. Particularly, there is a gap in the literature regarding structured approaches for utilizing process mining techniques in the validation assessments of ABS models~\citep{cabac2006analysis,bemthuis2023CoopIS}. Furthermore, while scholars have presented prototypes or artifact examples for illustrative purposes, the standardization and usability are limited~\citep{fleischmann2013subject}, which could potentially constrain accessibility and innovation in the field.

\subsection{Our contribution}
Methodologies employed in executing data mining or data science projects on agent-based systems have demonstrated efficacy. For instance, \citet{arroyo2010re} delineated a methodology for the integration of data mining within agent-based modeling. They clarify the conditions under which data mining proves advantageous, offering a description of each phase of the associated process. Indeed, some process mining project methodologies, like the one discussed by \citet{van2015pm}, provide guidance from a process mining vantage point. However, they are not specifically tailored for agent-based system use cases. We posit that a methodology should underscore the relationship between process mining and agent-based systems, drawing from the methodological frameworks extant in both the process mining and agent-based modeling domains. 

The discussions in this section underscore the need for a refined methodological approach to leverage data via process mining, contributing to the enhancement of ABS systems. However, while the utility of process mining in gleaning insights from agent-based systems is evident, there is limited methodological guidance pertaining to knowledge extraction from such systems via process mining techniques. In our work, we address this gap, proposing a method based on a well-known methodology for conducting data science projects~\citep{SCHROER2021526}, the CRISP-DM methodology, for the evaluation of ABS models through the use of process mining techniques. 

Using the CRISP-DM methodology offers several advantages. First, CRISP-DM has gained significant recognition due to its empirical achievements in addressing real-world challenges~\citep{SCHROER2021526}. Second, it outlines key stages that are crucial for data-driven problem-solving initiatives. This structured approach includes distinct phases and emphasizes the interactions between these phases, thereby providing guidelines and associated tasks~\citep{chapman2000crisp,wirth2000crisp}. Third, the modular nature of CRISP-DM highlights its adaptability, allowing for tailored modifications that align with specific contextual needs and facilitating integration with business practices, IT infrastructures, and other data mining methodologies~\citep{plotnikova2020adaptations}. 

While CRISP-DM’s flexibility is often lauded for its adaptability to diverse scenarios and projects, this flexibility can paradoxically lead to a lack of structure and clarity, particularly when implementation efforts are not meticulously planned or fail to adhere to guidelines~\citep{plotnikova2021adapting}. The iterative design of CRISP-DM anticipates multiple cycles through its phases. Yet, these iterations are frequently neglected or bypassed in practice, causing in incomplete or suboptimal results. This paradox, wherein CRISP-DM’s strength—its flexibility—induces a lack of structure and clarity, is not unique to CRISP-DM but is indicative of many data mining project methodologies. 

CRISP-DM, being the de-facto standard and an industry-independent process model for implementing data mining projects~\citep{SCHROER2021526}, is representative of numerous data science projects. Our methodology could serve as a preliminary step towards augmenting transparency in this seemingly paradoxical cycle, facilitated by the innovative, data-intensive process mining techniques. This has the potential to enhance our comprehension of how standardized data mining processes are extended and adapted in practice (e.g., identify perceived gaps in the CRISP-DM process and characterize how CRISP-DM is adapted to bridge these gaps). 

\section{Methodology for assessing agent-based simulation models}
\label{section:methodology}
Our methodology is based on the CRISP-DM methodology. The CRISP-DM methodology consists of six phases: business understanding, data understanding, data preparation, modeling, evaluation, and deployment~\citep{wirth2000crisp}. These phases collectively provide a structured and iterative approach to data mining projects. 

\begin{figure*}[ht!]
\centerline{\includegraphics[width=1.0\textwidth]{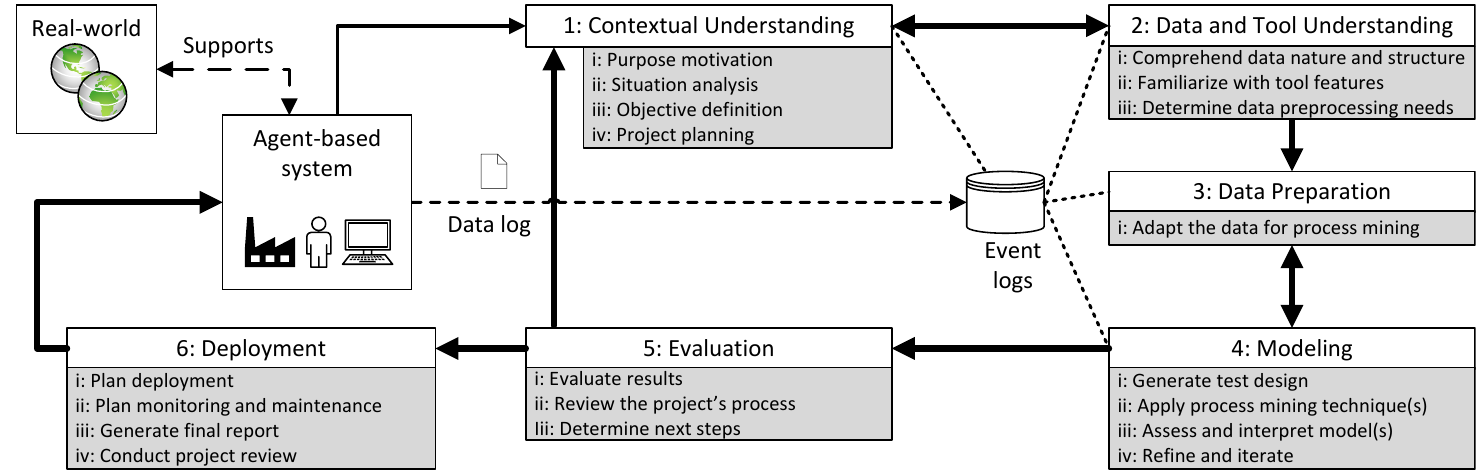}}
\caption{High-level overview of the methodology (adapted from \cite{chapman2000crisp} and \cite{wirth2000crisp})}
\label{fig:methodology}
\end{figure*}

Figure~\ref{fig:methodology} provides a high-level overview of our proposed methodology and reflects our adoption of the CRISP-DM methodology. Our method encompasses six distinct phases, each characterized by specific goals. Similar to CRISP-DM, our methodology operates iteratively. This means that the outcomes from one phase may necessitate revisiting previous phases. For instance, during the model evaluation phase, it may be necessary to revert to the data preparation phase to address issues that were previously unnoticed. We will discuss the methodology in more detail in the following. 

\subsection{Contextual understanding}
\label{subsection:contextual understanding}
The first phase involves understanding the broader context of the process mining project, which is intended for evaluating the agent-based system. The aim is to understand the overarching context and the reasons behind the use of process mining. This phase outlines the objectives and requirements of the entities involved. It is crucial to identify stakeholders, both individuals and groups, who have an interest in the project’s outcomes. The agent-based system can take several forms, such as real-world deployments or simulation models. 

The input for this phase is a detailed description of the agent-based system. The goals of this phase are to: (i) articulate the purpose or objective of the assessment; (ii) analyze the current situation, taking into account factors such as resources, requirements, assumptions, and stakeholders; (iii) define the objectives of process mining, including the criteria for success; and (iv) develop a project plan that includes a timeline, steps, and resource allocation. 

\subsection{Data and Tool Understanding}
\label{subsection:data understanding}
The second phase intends to establish an understanding of the data and tools. The inputs for this phase include raw data logs or traces from the agent-based system, process mining tools that are available along with their capabilities, and any previous analyses or models associated with the agent-based system. The goals for this phase are to: (i) understand the nature and structure of the data from the agent-based system; (ii) familiarize oneself with the features and limitations of the selected process mining tools or algorithms; and (iii) determine whether preprocessing or transformations are required for the data. 

Depending on the nature and quality of the data, specific tasks for understanding the data involve: collecting preliminary data from the agent-based system, characterizing the data, examining its attributes using visual or statistical methods, and ensuring the quality of the data. Tasks for understanding process mining tools can include: classifying the available tools and algorithms, assessing the compatibility and functionality of the tools, evaluating the usability of the user interface, reviewing the documentation and support, and exploring options for integration and interoperability. 

\subsection{Data Preparation}
\label{subsection:preparation}
This phase involves preparing the dataset(s) for use with process mining tools and algorithms. Inputs for this phase include (initial) data logs from the previous phase, identified data quality issues, inconsistencies, and data preprocessing tools or scripts. The objective of this phase is to convert the data into a format amenable to process mining. Depending on the quality and quantity of the data, tasks in this phase include selecting the subset of the dataset for use, cleaning, constructing, integrating, and formatting the data. For example, data can be structured in the XES format~\citep{verbeek2011xes}, a format widely supported by process mining tools, or an object-centric variant such as OCEL~\citep{ghahfarokhi2021ocel}. 

Nevertheless, please note that preparing datasets based on (snippets from) data from an agent-based system for use in process mining techniques is not a trivial task, especially when considering agent-related factors such as high dimensionality, hidden variables, emergent behaviors, temporal dynamics, and heterogeneity, as also highlighted by~\cite{tour2021agent} and~\cite{bemthuis2022discovering}. Furthermore, the quality and availability of data can depend on the volume of available log files or those that can be generated, the processing power at hand, the tools in use, and the nature of the agent-based model. This can also necessitate considerations such as the warm-up period, the cyclical nature of the system, and the presence of converging or diffusing behaviors. 

\subsection{Modeling}
\label{subsection:modeling}
The objectives of this phase are to: (i) generate a test design that details how the obtained process model(s) and insight(s) will be used to evaluate the ABS model; (ii) apply process mining techniques to extract process models and/or process insights (e.g., cycle times) from the ABS model data; (iii) assess and interpret the results of the model(s) based on domain knowledge, predefined success criteria, and the test design; and (iv) refine and iterate on the models as necessary to achieve a satisfactory model (i.e., one that can be worked with). Notably, the test design outlines how the results, derived from process mining techniques, are used for the evaluation of the ABS model.

\subsection{Evaluation}
\label{subsection:evaluation}
The evaluation phase aims to determine whether the outcome of the test design meets the project’s objectives and to outline the next steps. Specifically, the objectives are to: (i) examine the outcome of the test design (e.g., discovered process models) in the context of the project objectives, (ii) review the output of the overall project (e.g., were any steps overlooked or improperly executed?), and (iii) decide whether to proceed with the deployment, make adjustments, or halt the project. The output can include a detailed performance report of each obtained process model. Also, comparisons between models can be necessary, potentially involving users for model assessment, necessitating alignment between the complexity of a process model, and the user’s requirements when evaluating such a model. 

\subsection{Deployment}
\label{subsection:deployment}
The deployment phase involves using the obtained insights to enhance the agent-based system. The ultimate goal of an agent-based system is to improve real-life decision-making, and the deployment phase can contribute to achieving this goal. The complexities of the deployment phase depend on the nature of the project. The key objectives of this phase are to: (i) strategize the deployment of the insights gained, (ii) plan for monitoring and maintenance, (iii) generate the final report, and (iv) conduct a project review. 

\section{Illustrative case study}
\label{section:illustrative scenario}
In this section, we demonstrate the practical application of our proposed methodology using Schelling’s model of segregation, a renowned agent-based system. We aim to discuss the undertaken steps to provide a clear understanding of its operation within a suitable context. We selected Schelling’s model due to its seminal role in illustrating the emergence of collective patterns from individual preferences. Furthermore, its ubiquity and esteem in the field~\citep{ubarevivciene2024fifty} make it a promising test case for evaluating the efficacy and relevance of our methodology. 

Let us begin by providing a brief explanation of Schelling’s model. Since the 1970s, a substantial yet dispersed body of literature has helped shape the understanding of the societal issue of segregation~\citep{hatna2014combining,hegselmann2017}. Schelling's model illustrates how even mild individual biases, expressed through a level of tolerance, can result in pronounced segregation in communities~\citep{schelling1971dynamic}. We adopted the implementation of this agent-based system from~\cite{bemthuis2022discovering} and its subsequent refinements~\citep{bemthuis2023CoopIS,bemthuis2023EDOC}. The system was implemented using Python 3.6.9 using the AgentPy 0.1.5~\citep{foramitti2021agentpy} and PM4PY 2.7.2 libraries~\citep{berti2019process}. 

We refer to the work presented by~\cite{schelling1971dynamic} for an understanding of the basic principles of Schelling’s model. We will detail our further implementation (e.g., threshold levels) in the subsequent part of this section. 

\subsection{Contextual Understanding}
\label{subsection:case contextual understanding}

\subsubsection{Purpose motivation}
The goal of this project is to examine the impact of outlier behaviors on the validity of the ABS model. Investigating outliers in Schelling’s model helps in understanding the implications that arise from individual preferences. By focusing on outlier behaviors, we can gain insights into the dynamics of social systems and strategies to mitigate unintended societal segregation. Outliers can either challenge or reinforce a model’s credibility. Confirming certain atypical cases can enhance the simulation’s validity as they provide a more comprehensive reflection of reality. Conversely, identifying unrealistic behaviors can guide modifications to the simulation model, thereby enhancing its realism. The examination of rare events and extreme instances, in conjunction with the evaluation of these output values, can indicate potential inaccuracies in the model or questionable behaviors that do not faithfully represent reality \citep{carson2002model}. 

\subsubsection{Situation analysis}
Although we comprehend the implementation of Schelling’s model, as documented by~\cite{bemthuis2022discovering,bemthuis2023CoopIS,bemthuis2023EDOC}, we foresee the need for refinement in our current ABS model to more accurately mirror reality. This is a widely recognized challenge, highlighted by \cite{berg2010fast}, who contend that real-world segregation data frequently diverge from the predictions of the classic Schelling model, thus prompting their extension. Moreover, a bibliometric analysis conducted fifty years post the introduction of Schelling’s model of segregation revealed that these models serve as a source of inspiration for research rather than a subject for empirical testing~\citep{ubarevivciene2024fifty}. 

\subsubsection{Objective}
Our objective of employing process mining techniques is to identify outlier behaviors and evaluate modeling insights from them, which are instrumental in improving the realism of the simulation model. Process mining can be a potent tool for this purpose, with a specific focus on anomalous behaviors. 

\subsubsection{Refined objective and project plan}
We propose process mining as a tool for detecting potential weaknesses, investigating issues, conducting stress tests on the model, and identifying and rectifying problems. Consequently, we will demonstrate the application of process mining through one illustrative scenario, which will be discussed in the following sections. This article reflects the overall project plan. 

\subsection{Data and Tool Understanding}
\label{subsection:case data and tool understanding}
\subsubsection{Data nature and structure}
The selected ABS model can simulate a variety of scenarios, producing log files as output, as exemplified in Table~\ref{table:raw file}. The \textit{Step} attribute is related to the timestamp, and the \textit{StepCounter} represents an ordering counter within the timestamp. An agent transitions from a previous location \textit{PrevLoc} to a new location \textit{NewLoc} if the agent is not satisfied. The corresponding new \textit{Neighbors} and the number of agents from a similar group are recorded. The simulation model runs until it achieves a specified convergence level of happiness (of $100\%$) or completes a specified number of steps (100 in this case). The latter is chosen due to considerations such as computational constraints and the potential for models to converge without terminating. Preliminary data investigations, along with consideration of previous studies (e.g., \citet{bemthuis2023CoopIS,bemthuis2022discovering,bemthuis2023EDOC}), have ensured sufficient understanding of the nature and structure of the data. However, as will be discussed in the subsequent phase, the data requires some preprocessing. 

\begin{table*}[ht]
\caption{An excerpt of a raw data log}
\label{table:raw file}
\centering
\footnotesize
\begin{tabular*}{\hsize}{@{\extracolsep{\fill}}llllllll@{}}
\toprule
    EventNo & Step & StepCounter & AgentID & PrevLoc & NewLoc & Neighbors (AgentIDs)\\
    \midrule
        3359 & 4 & 53 & 2277 & (30, 17) & (25, 11) & [898, 541, 1564, 1914, 2091, 392]\\
        3360 & 4 & 54 & 2297 & (6, 16)  & (9, 40) & [1407, 2257, 1109, 411, 1087, 1940, 1269, 2182]\\
        3361 & 4 & 55 & 2326 & (19, 16) & (39, 15) & [1872, 886, 1239, 1711, 2229, 2273, 1747, 1034]\\
        3362 & 5 & 1  & 274 & (25, 30)  & (47, 8) & [1068, 28, 574, 542, 2143, 1461, 269, 1008]\\
        3363 & 5 & 2  & 297 & (39, 19)  & (3, 6) & [73, 2241, 556, 191, 328, 357]\\
        3364 & 5 & 3  & 439 & (46, 30)  & (21, 21) & [2077, 2076, 1048, 1195, 1372, 573, 1585, 1345]\\
    \bottomrule
\end{tabular*}
\end{table*}

Note that the data presented in Table~\ref{table:raw file} is illustrative; the ABS model can generate additional or different attributes, such as distance traversed or attributes linked to neighboring agents. Furthermore, attributes like \textit{Step} and \textit{StepCounter}, which represent logical sequences such as days or months, are adaptable examples within the model. Nevertheless, the purpose of this table is to give an idea of what can be produced and how the transformation from raw input data to an actual event log, which can be used by process mining discovery algorithms, is generated. 

\subsubsection{Tool features}
We select Disco as a process mining tool because (1) it can handle large event logs and complex models, (2) conversion and filtering are simplified, (3) performance metrics are presented directly and intuitively, and (4) it is user-friendly and offers intuitive drag-and-drop functionalities~\citep{gunther2012disco}. The discovery algorithm employed is the Fuzzy Miner, which provides a balance between simplifying complex processes for enhanced visualization and retaining enough detail to make the model informative, using two adjustable metrics: significance and correlation~\citep{gunther2007fuzzy}. 

\subsubsection{Data preprocessing}
To render the data compatible with the tool, the raw data log must be transformed into an event log file. This process is further discussed on Subsection~\ref{subsection:case data preparation}. Notably, this preparation data preprocessing step does not involve any additional tasks, such as filtering.

\subsection{Data Preparation}
\label{subsection:case data preparation}
The data set has been reformatted for easy integration with the tooling, while also adhering to the XES-format~\citep{verbeek2011xes}. Table~\ref{table:data file} provides an example of an event log derived from the raw data. The activities include three types: \textit{moveLocation} (indicating an agent's move from one location to another, implying discontent), \textit{changeHappy} (signifying a transition to a happy status for the agent), and \textit{changeUnHappy} (signifying a transition to an unhappy status for the agent). Similar to the work of~\cite{bemthuis2023EDOC}, we adopt a concatenated naming convention for analysis purposes, incorporating both the count of neighboring agents and those from a similar group: \textit{changeHappy\_X\_Y}, where \textit{X} represents the number of neighbors, and \textit{Y} denotes the count of neighbors from a similar group. We use the \textit{Step} number as a timestamp, interpreting it as a daily counter. Although the \textit{StepCounter} can be useful (e.g., for a more fine-grained analysis), we exclude it to maintain the demonstration's simplicity. Lastly, we only had access to the events that captured those agents who changed their status at least once (e.g., moved at least once). By focusing solely on those agents, we could reduce the complexity of the analysis, rendering the model easier to understand and interpret. 

\begin{table}[ht]
\caption{An excerpt of an event log generated from a raw data log}
\label{table:data file}
\centering
\footnotesize
\begin{tabular*}{0.8\linewidth}{@{\extracolsep{\fill}}llll@{}}
\toprule
Date & Activity & CaseID\\
    \midrule
17.10.2023 & \textit{move\_location} & 271\\
17.10.2023 & \textit{change\_happy\_2\_2} & 271\\
17.10.2023 & \textit{move\_location} & 280\\
18.10.2023 & \textit{move\_location} & 58\\
18.10.2023 & \textit{change\_unhappy\_7\_3} & 86\\
18.10.2023 & \textit{move\_location} & 137\\
18.10.2023 & \textit{move\_location} & 2\\
    \bottomrule
\end{tabular*}
\end{table}

\subsection{Modeling}
\label{subsection:case modeling}
We discuss the outcomes of the modeling phase after several iterations. 

\subsubsection{Test design}
\label{subsubsection:test design}
The test design describes the use of process models and insights for the evaluation of the ABS model. Our aim is to examine the impact of outlier behaviors on the validity of the ABS model. Therefore, we use the results obtained through process mining techniques as a surrogate measure for assessing the ABS model's fidelity to reality. 

Our illustrative case study focuses on the construct of face validity. Face validity involves soliciting expert opinions to ascertain the plausibility of a model and its behavior~\citep{sargent1992validation}. This process entails a subjective assessment of whether the process model appears to accurately represent reality~\citep{sargent2020verification,law2022build}. Face validity methods typically depend on human expertise, integrating expert evaluations, for instance, through systematic walkthroughs. While other validation techniques, such as statistical validation, hold importance, face validity often serves as the preliminary step in assessing a simulation model’s validity. This is particularly useful when there are limited or no formal methods available for validation~\citep{roungas2016towards}. Typically, after achieving a sufficient degree of validity based on a face validity check, more thorough and time-intensive checks are conducted. 

Several techniques exist within the face validity construct for evaluating ABS models, including graphical representation, immersive assessment, and sensitivity analysis~\citep{klugl2008validation}. Our test design emphasizes the use of a graphical display of a process model, which not only demonstrates relations but can also be enhanced with absolute or relative values in the realm of process models. An expert familiar with the application domain of segregation will assess the plausibility. 

The test design includes questions that the human assessor may deem plausible, not plausible, or requiring further investigation (i.e., cannot confirm or reject the observation’s plausibility). We provide a list of specific questions (refer to Subsection~\ref{subsubsection:assess and interpret the results}), based on a selection of indicators derived from the obtained process models. We regard any judgment made by the human expert as success criteria, as these could either reinforce or challenge the ABS model’s validity. 

While more systematic approaches could be implemented (refer to our discussion in Section~\ref{section:discussion}), we concentrate on presenting and discussing a subset of indicators that can be used for face validity assessment to demonstrate the effectiveness of our methodology. 

\subsubsection{Apply process mining techniques}
\label{subsubsection:apply process mining techniques}
We select a specific experimental setup of the ABS model to demonstrate a scenario for extracting insights about outlier behaviors, using process mining discovery techniques. In the setup, we set a uniform tolerance threshold of $0.55$ for all agents (i.e., the tolerated ratio of different neighbors before an agent decides to move), a grid size of $20$ by $20$ cells, and a density of $0.70$ (i.e., the ratio of occupied to empty cells). We applied the Fuzzy miner discovery technique on the event logs, using the $20024$ events (with $280$ cases and $46$ activities) as generated in log files. 

We will consider one illustrative scenario related to examining outlier behaviors based on indicators. We study a scenario, using as primary indicator on the process model case frequency, and as secondary indicator maximum number of repetitions within a case. The primary metric, \textit{case frequency}, is selected for its ability to quantify the recurrence of specific paths or activities within a process. High-frequency paths or activities are typically integral to the process, and their analysis can provide valuable insights into the emergence and evolution of segregation patterns. As a secondary metric, \textit{maximum number of repetitions}, provides insight into how often the process passed through a particular activity or path for the same case, instead of ignoring repetitions within the same case. This metric is particularly useful in identifying key paths or activities that, while not the most frequent, still account for a portion of the process instances. 

Additional metrics such as duration and sequence could also be informative. For example, duration could be relevant in processes where time efficiency is critical, while sequence could be vital in processes requiring adherence to a specific order of steps. However, in this illustrative scenario, our focus is on the aforementioned metrics. 

This scenario contributes to the project’s objectives by facilitating the identification of outliers through filtering at the level of displayed activities and paths. Contrary to initial impressions, focusing on these indicators might enhance outlier assessment. Establishing a ‘baseline’ allows for a comparative framework to identify outliers. Without this baseline, defining an outlier becomes challenging. Furthermore, large process models (see \ref{appendixA:complex process model} for an example of such a model) can be overwhelming and complex to analyze in their entirety. Scoping provides a manageable approach to analysis, simplifying the identification of outliers. 

We acknowledge that a more in-depth exploration of outliers based on complex process models could have been beneficial. However, the lack of structured methodologies for identifying outliers in process models derived from ABS models in literature poses a significant challenge~\citep{ghionna2008outlier}. We adopted a pragmatic filtering approach as an initial step (see next subsection). While this approach is not flawless, it enables us to manage the data complexity and concentrate on the key aspects of the methodology. It serves as a starting point for outlier identification and sets the stage for the adoption of more advanced methods. We recognize that we may miss some outliers that are only evident in the context of the full, unfiltered model. Nevertheless, we believe that the insights gained from the filtered model still provide valuable contributions to our understanding of the process. 

The specific questions posed to the human expert will be addressed subsequently. 

\subsubsection{Assess and interpret the results}
\label{subsubsection:assess and interpret the results}
We explore a selection of example indicators that either strengthen, refute, or question the validity of the ABS model. We consider each type of outcome as a ‘success’ criterion for the purpose of demonstrating our methodology. The indicators are assessed by a human expert. 

In this study, we investigate a scenario characterized by outlier behaviors. However, the process model resulting from the use of all event logs would be complex (refer to \ref{appendixA:complex process model} for an example), posing a challenge for face validity checks. To address this, we employ specific filtering techniques, confining our analysis to cases that meet the following criteria: 

\begin{itemize}
    \item The cases must intersect with the timeframe that begins one week after the start of the simulation. This criterion is implemented to mitigate the impact of the high frequency of movements observed in the initial state of the simulation. This initial state is random, leading to an anticipated high degree of segregation in the early stages. 
    \item We limit our analysis to cases that last less than $90$ days. This is based on the assumption that, beyond this period, an agent should have either found something or is presumed to be relocating outside our considered grid. 
    \item Our analysis is further restricted to cases with no more than $25$ events. This restriction is based on the assumption that cases with an event count exceeding this limit may lack realism. However, still a substantial number of events are deemed necessary for the development of a process model with reasonable coverage. 
\end{itemize}

Applying these filters results in the inclusion of $12$ cases ($4\%$) and $180$ events ($\approx 1\%$). The process model, obtained using the Fuzzy miner, is shown in Figure~\ref{fig:example b}. Using the insights from case frequency and maximum number of repetitions, and guided by the color scheme employed in the process model, we have consequently proposed a set of questions related to the indicators. These questions are designed to assist a human expert in the assessment of face validity. The evaluation criteria per observation are based on work reported by \cite{bemthuis2023EDOC}, and based on the questions: 

\begin{figure}[htb!]
  \centering
  \includegraphics[angle=90,height=0.88\textheight]{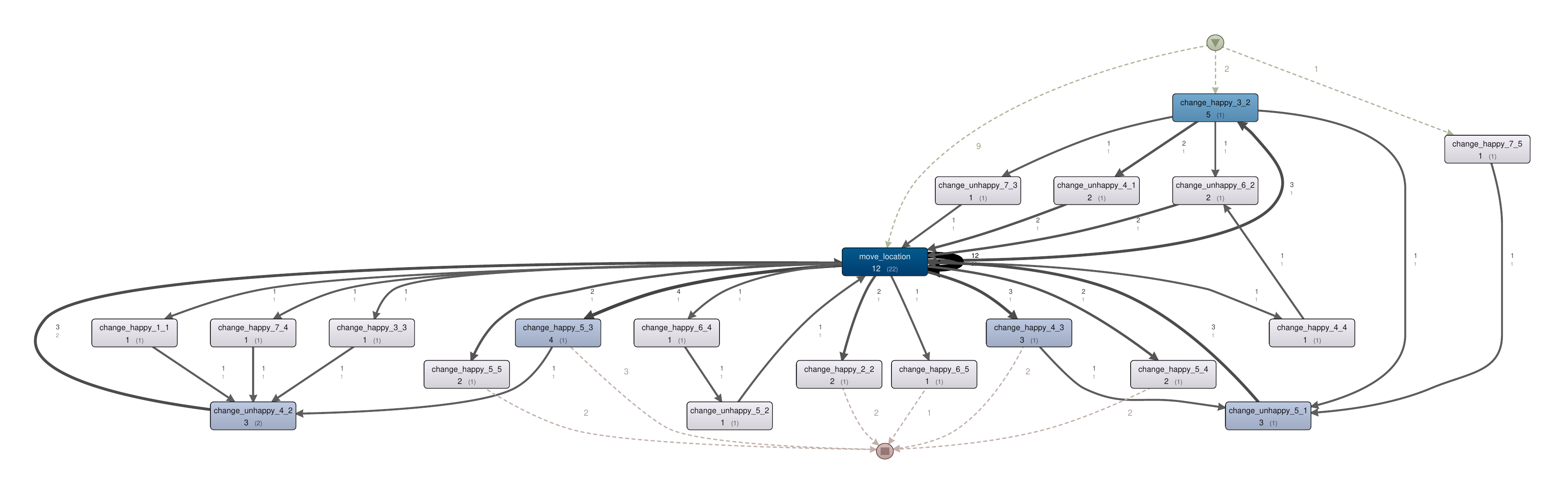}
  \caption{Process model abstraction with $100\%$ activities and $100\%$ paths (primary indicator: case frequency; secondary indicator: maximum number of repetitions).}
  \label{fig:example b}
\end{figure}

\begin{enumerate}
    \item Does the obtained performance indicator value accurately reflect the real-world system being modeled? 
    \item Given an overall population size of $280$ agents, can the obtained performance indicator value be considered a plausible representation? 
\end{enumerate}

While the first question directly addresses the observation, the second question allows us to interpret the results within a broader context. This interpretation takes into account the pre-filtered population size, using the filtered outcome as a representative value. The questions, including the assessment as done by a human expert familiar with segregation, are presented in Table~\ref{table:questions}. 

\begin{table*}[htb!]
\caption{A selection of results of the face validity assessment as judged by an expert}
\label{table:questions}
\centering
\footnotesize
\begin{tabular*}{0.95\linewidth}{p{0.01\linewidth}p{0.38\linewidth}p{0.12\linewidth}p{0.16\linewidth}p{0.16\linewidth}}
\toprule
\# & Activity/path & Observation* & Question 1 & Question 2\\
    \midrule
1 & ${move\_location}$ & CF=$12$ ($100\%$)& not plausible & plausible\\
2 & ${move\_location}$ & MNR=$22$ & not plausible & further investigation\\
3 & ${move\_location} \rightarrow {move\_location}$ & CF=12 ($100\%$) & not plausible & not plausible\\
4 & ${change\_happy\_3\_2}$ & CF=$5$ ($42\%$) & further investigation & plausible\\
5 & ${change\_happy\_3\_2}$ & MNR=$1$ & plausible & plausible\\
6 & ${move\_location} \rightarrow {change\_happy\_5\_3}$ & CF=$4$ ($33\%$) & plausible & plausible\\
7 & ${change\_unhappy\_4\_2}$ & CF=$3$ ($25\%$) & plausible & plausible\\
8 & ${change\_unhappy\_4\_2}$ & MNR=$2$ & further investigation & plausible\\
9 & ${change\_happy\_7\_4} \rightarrow {change\_unhappy\_4\_2}$ & CF=$1$ ($8\%$) & further investigation & plausible\\
    \bottomrule
\multicolumn{5}{l}{*: CF = case frequency; MNR = maximum number of repetitions.}\\
\end{tabular*}
\end{table*}

We now present a brief discussion of some observations, referencing the numbering in Table~\ref{table:questions}. The expert deemed the first observation not plausible, arguing that it is unrealistic for all households to have moved at least once, given that many households are generally content with their housing situation. This observation may be a consequence of data selection in the methodology (see Subsection~\ref{subsection:case data preparation}). However, considering the broader context of that this could reflect $12$ out of $280$ cases, the expert conceded that this number might be plausible. For the second observation, the expert posited that the number of repetitions was too high to realistically reflect reality. Yet, in the context of the entire population (question 2), the expert admitted a lack of recall regarding specific numbers, suggesting a need for further investigation. As for the third observation, the expert found it surprising and not plausible that each agent that moved once was immediately followed by another movement activity, even when considering the broader context of the entire population (question 2). The remaining observations, as presented in the table, are left for the reader’s consideration. 

\subsubsection{Refine and iterate}
For the purpose of this study, we limit our focus to the current findings. These results, while insightful, represent a snapshot in an ongoing process of exploration and development of the ABS model. Nevertheless, based on the observations from the expert, some suggestions for refinement were proposed. Firstly, the inclusion of a broader spectrum of agent behaviors, as well as a more diverse array of agents (for instance, stationary agents that may nonetheless experience environmental changes), could provide additional insights. Secondly, the expert evaluated the observations independently, treating each as a separate query. However, it might be beneficial to consider multiple performance indicators simultaneously to form a more comprehensive judgement about specific observations. Future research may expand on these findings, reflecting the iterative nature. 

\subsection{Evaluation}
\label{subsection:case evaluation}

\subsubsection{Evaluate results}
For both questions, we report all three assessment outcomes, reflecting a range of validity judgments. The most frequent assessment for question 2 is “Plausible”, implying that most observations are deemed potentially valid upon preliminary review. The distribution of assessments for question 1 is more balanced, suggesting a diverse initial appraisal of the model’s validity. Observations 2, 3, 4, 8, and 9 exhibit discrepancies between the assessments for questions 1 and 2, indicating possible areas for further investigation or divergent views on these activities’ validity. 

In summary, not all outcomes contribute positively to the realism of the ABS model. Considering question 1, only three observations were classified as plausible, three necessitate further scrutiny, and three were deemed not plausible. Despite these mixed results, we identified indicators that either affirm or necessitate further investigation to properly evaluate their validity. While this aligns with our project’s success criteria, it casts doubt on the ABS model’s validity (which is not detrimental to our study’s objectives). 

Please note that in our illustrative example, we engaged a single human assessor for evaluation, although the inclusion of a broader panel of experts acquainted with Schelling's model and/or segregation dynamics could have enriched the assessment. Our assessor relied on tacit knowledge for judgment, whereas an explicit, qualitative evaluation might have yielded more insightful results. Engaging multiple experts could provide a more comprehensive perspective, but it also introduces the risk of conflicting outcomes, potentially obstructing the swift execution of a face validity check. Therefore, caution must be exercised when conducting face validity exercises, especially when the anticipated output is already known, rendering the model redundant~\citep{law2022build}. 

Furthermore, we only present a subset of the formulated questions in this study. These questions span various activities, derived from the color schemes, and flows, derived from the thickness of the arrows. Our aim is not to provide an exhaustive list of all possible questions, but rather to offer a selection of potential questions pertinent to this study. An exhaustive list, while comprehensive, may not be practical for a face validity check. 

\subsubsection{Review the project's process}
Recall that the goal of this illustrative case study is to examine the impact of outlier behaviors on the validity of the ABS model. The application of process mining techniques has yielded several indicators that either refute, question, or reinforce the ABS model’s validity. However, it is important to note that our illustrative case study, constrained by time and resources, only identifies a limited number of these indicators from the viewpoint of a single human evaluator. This necessitates a more comprehensive exploration. 

A variety of process mining techniques are at our disposal, which can be used to identify and evaluate outliers based on factors such as duration, frequency, or sequence. However, In relation to outliers, process mining practices typically embrace straightforward and practical solutions~\citep{ghionna2008outlier}. Our approach aligns with this, but we also advocate for the incorporation of self-tuning techniques and methods that generate hypotheses for pinpointing critical points that can result in invalid (outlier) behaviors. 

In addition, while our examples underscore the application of process mining techniques, it is also worth considering quality metrics for a discovered process model, such as fitness, precision, and generalization. This approach ensures a more thorough and robust evaluation of the model’s validity. 

Lastly, we have presented the results chronologically, even though our methodology incorporates iterative feedback loops, facilitating the continuous refinement and improvement of the project. However, the iterative nature of the methodology and its feedback loops can be difficult to represent in a linear, written format, potentially hindering the reader’s understanding of the process and the results. Time and resource constraints are among other challenges faced.

\subsubsection{Determine next steps}
We have presented the results of a preliminary evaluation, which aligns with a face validity check. Face validity, an informal validation technique, is often used for a quick review of simulation model outcomes to assess their plausibility. Despite its informal nature and the criticism it receives from validity theorists~\citep{royal2016face}, face validity offers a means to evaluate the results of an ABS model~\citep{klugl2008validation}. Therefore, while the concept of face validity is a subject of scholarly debate, it can still aid in interpreting the model’s validity. 

The results discussed herein contribute to the validation process of the chosen ABS model. Schelling’s models, despite their relative simplicity and comprehensibility, pose complex testing challenges. This complexity is further underscored by~\cite{ubarevivciene2024fifty}, whose claim is grounded in a bibliometric analysis that covers fifty years of research on Schelling’s model. 

In our study, we have made contributions by providing indicators that aid in conducting face validity. However, we advocate for additional testing and iterations using the proposed methodology, which could pave the way for a more profound understanding of spatial segregation. 

In the wider context of ABS model validation, it is important to recognize that theories can be challenged and disproven, but they are seldom definitively validated. Similarly, when a model is deemed valid, certain aspects may have been overlooked, or new insights about the real-world counterpart may indicate that the simulation model is no longer valid. For instance, an outlier in a disease spreading simulation model (e.g., a ‘super-spreader’) might initially be deemed as not accurately representing reality. However, as time progresses, such a scenario could indeed occur in a real-world system~\citep{bemthuis2023EDOC}. Therefore, a model’s ability to accurately predict the performance of the corresponding real system under specific conditions does not necessarily guarantee the universal accuracy of its predictions~\citep{murray2015testing}. 

Thus, while our methodology and application to Schelling's model represent an advancement, there is an evident need for the development of more rigorous and procedural approaches to validate simulation model. Such development could ensure the accuracy of the models and expand their applicability across a broader range of scenarios. Future research should concentrate on addressing these challenges to further propel the field of ABS model validation. 

\subsection{Deployment}
The actual deployment falls outside the purview of this illustrative scenario, primarily due to time and resource constraints. Real-world case studies based on Schelling's model often encompass complexities that transcend mere technical challenges, incorporating organizational, political, and economic factors. Additionally, our investigation focused on assessment rather than the actual implementation of improvement suggestions. 

\section{Discussion}
\label{section:discussion}
In this section, we discuss the relevance of this study by reflecting on its theoretical and practical limitations and implications, thereby emphasizing our contributions. This research has several limitations and implications, which we will discuss now. 

First, we address the assessment of agent-based systems, but we provide only a limited overview of the available assessment methods pertaining to specific properties of agent-based systems. Multi-agent systems, encompass numerous heterogeneous agents with distinct characteristics. This diversity presents considerable challenges in establishing specific evaluation criteria. Evaluations may explore various aspects such as agent communication, collaboration mechanisms, coordination strategies, and self-learning capabilities. Each of these functionalities warrants refinement within our proposed methodology. For instance, the development of specialized process mining discovery algorithms is crucial for extracting knowledge that targets the properties of an agent system. 

Second, our methodology can be further developed to enhance the capabilities of the agent-based system being analyzed, whether they are virtual representations of reality or real-world systems operating under agent-based paradigms. Once the assessment is complete (or at least obtained evidence related to the ABS model's validity), it would be a missed opportunity not to utilize the gained insights. These insights can be used for developing a more accurate and enhanced ABS model. 

Third, our proposed method is not intended as a prescriptive mandate but as a foundational guide. The objective is not to deliver an exhaustive or definitive approach, but to furnish newcomers in this domain with an initial framework. We provided a list of proposed tasks, the execution of which may depend on various factors such as the availability and quality of data, as well as resource constraints like computational power. Therefore, our tasks should be viewed as suggestions rather than requirements. We expect that as scholars and practitioners delve deeper into the field, they will inevitably refine and augment this method based on their distinct needs and insights. Furthermore, given the widespread acknowledgment and adoption of the CRISP-DM methodology, the integration of our method should be relatively seamless. Those acquainted with CRISP-DM will readily adapt our steps, merging them into their established workflows and possibly crafting extensions for enhanced clarity and depth. We therefore advise the application of specific techniques in accordance with the CRISP-DM methodology. 

Fourth, it would be beneficial to provide guidance on how to leverage the plethora of applications from the process mining discipline for the agent-based community, and vice versa. While design methodologies are prevalent in both the process mining field (e.g.,~\citealp{van2015pm}) and the agent-based modeling field (e.g.,~\citealp{zambonelli2003developing}), a gap persists in the literature concerning the correlation between agent-based functionalities and process mining capabilities. Our methodology lays the groundwork for users to further cultivate these connections. Nevertheless, additional formalization is needed. For instance, the employment of domain ontologies, data specifications, and design protocols, such as the Overview, Design concepts, and Details (ODD) protocol~\citep{grimm2020odd}, could prove beneficial. 

Fifth, the (nontrivial) generation of event logs during early stages of simulation model construction is crucial. This process allows for preliminary assessments of an ABS model’s validity. Early identification of significant discrepancies between the simulation model and reality can mitigate the risk of inappropriate simulation use~\citep{pace2003verification}. Moreover, it can avert error propagation, which could necessitate considerable rework if identified later in the process~\citep{sargent2020verification}. Consequently, our endeavors, particularly the case study that concentrates on the face validity construct typically utilized during the early building and validation stages, are instrumental in ensuring the model's precision and reliability during the initial stages of simulation model construction. Thus, the sooner and more frequently input is solicited from individuals involved in conducting face validity checks, the more it may contribute to face validity~\citep{lucko2010research}. This, in turn, enhances the confidence of users and decision-makers. Nevertheless, although face validity is often considered informal and inconsistent, it pertains to an artifact that can at least be evaluated~\citep{klugl2008validation}. 

Sixth, we discuss the construct of validation within the realm of ABS models. However, we should be cautious and critical about the extent to which process mining is used as a surrogate for ultimately producing a precise and credible simulation model. Analysts in agent-based modeling do not only examine inputs such as initial conditions and model specifications, but they also consider the history or paths taken to reach the final state. They frequently assess simulations that evolve over time using a variety of techniques, including examining variables beyond those measurable in experiments~\citep{bemthuis2023CoopIS}. Validation of ABS models can be accomplished through a multitude of techniques (see, for example,~\citealp{kleijnen1995verification,klugl2008validation,hora2015review}). Process mining can be viewed as just one of these techniques, and potential new biases introduced by the use of process mining techniques should also be taken into account. 

Seventh, our methodology is grounded in the CRISP-DM methodology and adjusted to be used to evaluate ABS models based on process mining techniques. We apply process mining techniques solely to the output from an ABS model. This restricts us from conducting ‘meta-level’ analyses, such as scrutinizing the actual execution of the methodology. Furthermore, indicators derived not only from the obtained process models (e.g., number of places, transitions, and edges, etc.) but also those associated with the process mining discovery algorithms used (e.g., computational speed), along with other potential insights obtained through other process mining techniques (e.g., conformance checking), could enrich our understanding of an ABS model's performance. Thus, process mining can aid in upholding high standards in ABS model construction and adhering to best practices of transparency and rigor for replicability, as also called for by~\cite{squazzoni2020computational}. 

\section{Conclusions and future work}
\label{section:conclusions}
In this paper, we proposed a methodology for assessing ABS models using process mining techniques. The methodology draws from the CRISP-DM methodology and is demonstrated using an illustrative scenario based on Schelling’s model of segregation. By integrating process mining techniques into stages of the CRISP-DM methodology, our approach enables a structured analysis of ABS models. Our illustrative scenario using Schelling’s model of segregation demonstrates that we can effectively assess ABS model outcomes, offering insights into the system’s behavioral patterns (based on outlier behaviors). This allows us to identify key factors influencing the model’s behavior. Our step-by-step approach has potential for application across a wide range of ABS models and domains. 

Further research is necessary to establish the comprehensiveness of our methodology and to explore potential extensions. We recommend further investigations and experiments with more complex, real-life case studies. Additionally, it is important to determine the extent to which our methodology can be applied to specific ABS model validation but also verification methods. Process mining techniques can facilitate interaction with simulation modelers and assessors, ultimately enhancing the accuracy and realism of ABS models. This is not limited to the use of process mining techniques alone; integration with other data mining and machine learning techniques can further enhance the evaluation of ABS model behavior and produce human-interpretable results. Future research could also focus on the automation of our methodology's application, as well as exploring the use of process mining to enhance not only the validity of the ABS model but also the execution of the methodology itself. A promising avenue for future research is the application of object-centric process mining, which could offer a novel perspective for analyzing ABS models. 

\appendix

\section{Example of a complex process model}
\label{appendixA:complex process model}
\begin{figure}[ht!]
    \centering
    \includegraphics[angle=90,height=0.88\textheight]{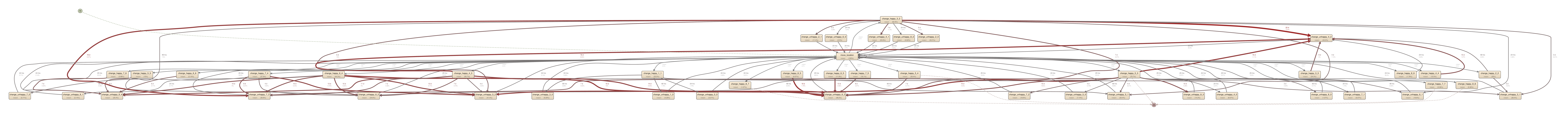}
    \caption{Example of an obtained process model abstraction with $100\%$ activities and $15\%$ paths (primary indicator: maximum duration; secondary indicator: case coverage)}
    \label{fig:example of large model}
\end{figure}

Figure~\ref{fig:example of large model} depicts an example of a process model containing many activities and paths. 

\section*{Acknowledgements}
This project has received financial support from the ECOLOGIC project, which was funded by the Dutch Ministry of Infrastructure and Water Management and TKI Dinalog (application number 5000006252 and case number 31192090). The authors would like to express their gratitude to Faiza Bukhsh for the productive discussions. 

\bibliographystyle{elsarticle-harv} 
\bibliography{bibdatabase}

\end{document}